\documentclass[12pt,a4paper]{article}
\usepackage[cp1251]{inputenc}
\usepackage[english]{babel}
\usepackage{graphicx}
\usepackage{amsmath}
\date{} \lccode`\-=`\-
\title{DARBOUX TRANSFORMATION OF THE GREEN
FUNCTION FOR THE DIRAC EQUATION WITH THE GENERALIZED POTENTIAL}
\author{Ekaterina Pozdeeva\\
 {\it Department of Quantum Field Theory,}\\
 {\it Tomsk State University, 36 Lenin Avenue}\\
{\it Tomsk, 634050, Russia}\\
{\it ekatpozdeeva@mail.ru} \\\\
Received 21 July 2007\\
 Revised 17 September 2007\\ Managing editor V.
A. RUBAKOV}
\begin{document}
\selectlanguage{english} \maketitle

\begin{abstract}
We consider the Darboux transformation of the Green functions of the
 regular boundary problem of the one-dimensional stationary Dirac equation.
 We obtained the Green functions  of the transformed Dirac equation with the
initial regular boundary conditions. We also construct the formula
for the unabridged trace of the difference of the transformed  and
the initial Green functions of the regular boundary  problem of the
one-dimensional stationary Dirac equation. We illustrate our
findings by the consideration of the Darboux transformation
 for the Green function of the free particle Dirac equation on an
 interval.\\

\end{abstract}
\section{Introduction}

There has been great interest in using there Darboux transformation
\cite{Darbu, Rosu} for the analysis of physical systems \cite{Axel1,
Axel2, Song2} and for finding new solvable systems
\cite{Matv,Suzko,Pozdeeva}. It has been shown that the
transformation method is useful in finding soliton solutions of the
integrable systems \cite{Park,Abl} and constructing supersymmetric
quantum mechanical systems
\cite{Witten,BAGROVECHAR}.\footnote{Recently a special issue of {\it
Journal of Physics,} A {\bf34}  was devoted to research work in
supersymmetric quantum mechanics (SUSY QM).}  It is well known that
SUSY QM is basically equivalent to the Darboux transformation and
the factorization properties of the Schr\"o\-din\-ger equation
\cite{Sukumar1985, BagrovSamsonov, Gomez}.  The Darboux
transformation of the one-dimensional stationary Dirac equation is
equivalent to the underlying quadratic supersymmetry  and the
factorization properties \cite{annphys2003v305p151,Eurjphys,BAGROV}.

Despite of the growing number of papers in this field many questions
still remain open and require further
 study. In particular, the author are aware of only several papers
 \cite{Sukumar, Samsonov, Pupasov} devoted to the SUSY transformations at the level of
 the  Green functions.  In \cite{Sukumar} was obtained  the integral relation between the Green
 functions for two SUSY partner Hamiltonians of the one-dimensional Schr\"o\-din\-ger equation  with discrete spectra.
 In \cite{Samsonov, Pupasov} the integral relation between
the Green functions corresponding two SUSY partner Hamiltonians of
the one-dimensional Schr\"o\-din\-ger equation  is generated to the
case of continuous spectrum. The exact Green function of the
time-dependent Schr\"o\-din\-ger equation was obtained in
\cite{Anderson, Axel4}.

An interesting open problem is to find the analogous results for the
Dirac equation. In this paper, we construct the Darboux
transformation of the Green function for the regular boundary
problem of the one-dimensional stationary Dirac equation with a
generalized form of the potential and obtain formulas for the
unabridged trace of the difference of the modified and initial Green
functions. The rest of the paper is organized as follows. In Section
2 we construct a Green function for the initial regular boundary
problem of the one-dimensional Dirac equation with the generalized
form of the potential. In Section 3 we consider the Green function
for the transformed regular boundary problem of the one-dimensional
Dirac equation. We construct the Darboux transformation  of the
Green function. In Section 4 we obtain formulas for the unabridged
trace of the difference of the modified and the initial Green
functions and consider the spectral representation of the
corresponding unabridged trace. In Conclusion the summarize our
results and speculate about some perspective.
\section{Green function of the one-dimensional Dirac equation}

 The Green function of the one-dimensional Dirac
equation
\begin{equation}
\label{1} (h_0(x)-E)\psi(x)=0
\end{equation}
with the Dirac Hamiltonian of the form
$h_0=i\sigma_2\partial_x+V(x)$ \cite{Deber} is needed for obtaining
the solution
\begin{equation}
\label{thesolution}
\Phi(x,E)=\int_a^bG_0(x,y,E)F(y)dy\end{equation}of the following
inhomogeneous equation
\begin{equation}
 \label{2}
 (h_0(x)-E)\Phi(x,E)=F(x),
\end{equation}
where  $a$ and $b$ are the endpoints of an closeg interval.

Earlier this method allowed the authors of the paper \cite{levitan}
to obtain the Green function of the one-dimensional Dirac equation
with the potential
$$U_0=\left(\begin{array}{cc}\ p(x)&0\\0&q(x)
\end{array}\right).$$

 In the present paper, we will obtain the Green function of the Dirac equation
 with the generalized form \cite{thaler} of the potential
 \begin{equation}\label{3}
    V_0(x)=\omega(x)I+(m+S(x))\sigma_3+q(x)\sigma_1,
\end{equation}
where $\omega(x)$, $S(x)$ and $q(x)$ are real functions of $x$, $m$
is the mass of a particle, $\sigma_1$, $\sigma_3$ are usual Pauli
matrices. The generalized potential is the self-adjoint potential.

 We consider the following boundary conditions for the components $\Phi_1(x,E)$, $\Phi_2(x,E)$
  of the  solution (\ref{thesolution}) to
 (\ref{2}):
 \begin{eqnarray}
 \label{problem}
  \label{s1}
\Phi_1(a,E)\sin(\alpha)+\Phi_2(a,E)\cos(\alpha)=0,
\end{eqnarray}
\begin{eqnarray}
 \label{s2}
\Phi_1(b,E)\sin(\beta)+\Phi_2(b,E)\cos(\beta)=0.
\end{eqnarray}

We suppose that the solutions $\psi(x)$, $\varphi(x)$ of the Dirac
equation with the generalized form of the potential obey the
following conditions:
\begin{eqnarray}
 \label{5}
\varphi(a,E)=\left(\begin{array}{c}
   \cos\alpha \\
  -\sin\alpha\\
\end{array}\right),&
&\psi(b,E)=\left(\begin{array}{c}
   \cos\beta\\
  -\sin\beta\\
\end{array}\right).
\end{eqnarray}

Let us construct the matrix
\begin{eqnarray} \label{G}
G_0(x,y,E)=\frac{1}{W(E)}\left\{\begin{array}{cc}\
\psi(x,E)  \varphi ^T(y,E), & y\leq x,\\
 \varphi(x,E)\psi^T (y,E), &
x<y,\end{array}\right.
\end{eqnarray}
where $W(E)=W\{\varphi(x,E),\psi(x,E)\}=const$ is the Wronskian of
the two functions $\psi(x)$ and $\varphi(x)$.

One can reality examine that the vector-function $\Phi(x,E)$ is the
solution to the inhomogeneous equation (\ref{2}) and therefore the
matrix (\ref{G}) is the Green function of the regular boundary
problem of the one-dimensional stationary Dirac equation with the
generalized form of the potential (\ref{3}).

Let us construct the spinor $y(x,E)=\left(
\begin{array}{c}
  y_1(x,E) \\
  y_2(x,E) \\
\end{array}
\right)$
\begin{equation}
 \label{y}
  y(x,E)=\int^{a}_{b}G_0(x,y,E)F(y)dy.
 \end{equation}

In detail,  Eq.  (\ref{G})  becomes
\begin{equation}\label{GGG}
    G_0(x,y,E)=\frac{1}{W(E)}\left\{
\begin{array}{cc}
  \left(
\begin{array}{cc}
  \varphi_1(x,E)\psi_1(y,E) & \varphi_1(x,E)\psi_2(y,E) \\
  \varphi_2(x,E)\psi_1(y,E) & \varphi_2(x,E)\psi_2(y,E) \\
\end{array}
\right), & x<y \\
 \left(
\begin{array}{cc}
 \varphi_1(y,E)\psi_1(x,E) & \varphi_2(y,E)\psi_1(x,E) \\
  \varphi_1(y,E)\psi_2(x,E) & \varphi_2(y,E)\psi_2(x,E) \\
\end{array}
\right),  & y\leq x. \\
\end{array}
\right.
\end{equation}

We write down a detailed form of the components of the spinor
\begin{equation}\label{GFF}
G_0(x,y,E)F(y)=\left(
\begin{array}{c}
 \Gamma_1 \\
 \Gamma_2 \\
\end{array}
\right),\end{equation}
\begin{equation}\label{GF1}
    \Gamma_1=\frac{1}{W(E)}\left\{
\begin{array}{cc}
  \varphi_1(x,E)\psi_1(y,E)F_1(y)+\varphi_1(x,E)\psi_2(y,E)F_2(y), & x<y,
  \\ \nonumber
  \varphi_1(y,E)\psi_1(x,E)F_1(y)+\varphi_2(y,E)\psi_1(x,E)F_2(y),  & y\leq x,\nonumber
\end{array}
\right.
\end{equation}
\begin{equation}\label{GF2}
   \Gamma_2=\frac{1}{W(E)}\left\{
\begin{array}{cc}
     \varphi_2(x,E)\psi_1(y,E)F_1(y)+\varphi_2(x,E)\psi_2(y,E)F_2(y), & x<y
    \\ \nonumber
    \varphi_1(y,E)\psi_2(x,E)F_1(y)+\varphi_2(y,E)\psi_2(x,E)F_2(y),  & y\leq x. \nonumber
\end{array}
\right.
\end{equation}

From these equations, taking into consideration  (\ref{GFF}) and
$$\psi_1(y,E)F_1(y)+\psi_2(y,E)F_2(y)=\psi^T(y,E)F(y),$$
$$\psi_2(y,E)F_1(y)+\varphi_2(y,E)F_2(y)=\varphi^T(y,E)F(y),$$ we
find the components of the spinor \eqref{y} $y_1(x,E)$, $y_2(x,E)$.

We consider the components of the spinor \eqref{y} in the case
$y\leq x$, then $y\in[a,x]$, after that in the case  $y>x$, when
$y\in(x,b]$, and as a result, we find  the components of the spinor
\eqref{y}  $y_1(x,E)$, $y_2(x,E)$ in the general case:
\begin{align}
  y_1(x,E)=\frac{1}{W(E)}\{&\psi_1(x,E)\int^{x}_{a}\varphi^T(y,E)F(y)dy+ \nonumber\\
   &+\varphi_1(x,E)\int^{b}_{x}\psi^T(y,E)F(y)dy\}\label{12},
\end{align}
\begin{align}
  y_2(x,E)=\frac{1}{W(E)}\{&\psi_2(x,E)\int^{x}_{a}\varphi^T(y,E)F(y)dy+ \nonumber\\
   &+\varphi_2(x,E)\int^{b}_{x} \psi^T(y,E)F(y)dy\}. \label{14}
\end{align}

Now we check that  the vector-function $y(x,E)$ is the solution of
the inhomogeneous equation (\ref{2})
\begin{equation}
 \label{16}
 \partial_xy_2(x,E)-(E-S(x)-m-\omega(x))y_1(x,E)+q(x)y_2(x,E)=F_1(x),
\end{equation}
\begin{equation}
 \label{17}
 \partial_xy_1(x,E)+(E+S(x)+m-\omega(x))y_2(x,E)-q(x)y_1(x,E)=-F_2(x),
\end{equation}
where  $F_1(y)$, $F_2(y)$ are the components of the spinor $F(y)$.

Let us show that equality (\ref{17}) is true. For that  we
differentiate  $y_1(x,E)$:
\begin{align}
\partial_xy_1(x,E)&=\frac{1}{W(E)}\{\partial_x\psi_1(x,E)\int_a^x\varphi^T(y,E)F(y)dy\nonumber\\
 &+\partial_x\varphi_1(x,E)\int_x^b\psi^T(y,E)F(y)dy\}\nonumber\\
&-\varphi_1(x,E)\{\psi_1(x,E)F_1(x)+\psi_2(x,E)F_2(x)\}\nonumber\\
&+\psi_1(x,E)\{\varphi_1(x,E)F_1(x)+\varphi_2(x,E)F_2(x)\}.\label{127777}
\end{align}
Since  the spinors $\psi(x,E)$ and $\varphi(x,E)$ are the solutions
to the Dirac system of equations, the following equalities are
correct:
\begin{eqnarray}\label{12717}
   \partial_x\psi_1(x,E)&=&q(x)\psi_1(x,E)-(S(x)+m-\omega(x)+E)\psi_2(x,E),
\end{eqnarray}
\begin{eqnarray}\label{1272}
  \partial_x\varphi_1(x,E)&=&q(x)\varphi_1(x,E)-(S(x)+m-\omega(x)+E)\varphi_2(x,E).
\end{eqnarray}
Substituting equalities (\ref{12717}), (\ref{1272}) into
(\ref{127777}), we obtain
\begin{align}\label{128}
    \partial_xy_1(x,E)&=\frac{1}{W(E)}[q(x)\{\psi_1(x,E)\int_a^x\varphi^T(y,E)F(y)dy\nonumber\\
    &+\varphi_1(x,E)\int_x^b\psi^T(y,E)F(y)dy\}\nonumber\\
    &-(S(x)+m-\omega(x)+E)\{\psi_2(x,E)\int_a^x\varphi^T(y,E)F(y)dy\nonumber\\
    &+\varphi_2(x,E)\int_x^b\psi^T(y,E)F(y)dy\}]-F_2(x).
\end{align}
From  (\ref{12}), (\ref{14}) for $y_1(x,E)$, $y_2(x,E)$ we get
\begin{equation}\label{129}
    \partial_xy_1(x,E)=q(x)y_1(x,E)-\{S(x)+m-\omega(x)+E\}y_2(x)-F_2(x).
\end{equation}
Therefore, (\ref{17}) is correct. Similarly, equality (\ref{16}) can
be proven. The spinor-function $\Phi(x,E)=y(x,E)$ is the solution to
(\ref{2}) with  the generalized form of the potential.

Now we demonstrate realization of  the regular boundary conditions
(\ref{s1}), (\ref{s2}) for  spinor-function $\Phi(x,E)=y(x,E)$
 \begin{eqnarray}
 \label{problem}
  \label{sy1}
y_1(a,E)\sin(\alpha)+y_2(a,E)\cos(\alpha)=0,
\end{eqnarray}
\begin{eqnarray}
 \label{sy2}
y_1(b,E)\sin(\beta)+y_2(b,E)\cos(\beta)=0,
\end{eqnarray}
where $y_1(x,E)$, $y_2(x,E)$ are the components of the spinor
$y(x,E)$.
 We compute (\ref{12}),
(\ref{14}) when  $x=a$, $x=b$:
\begin{align}
  y_1(a,E)=\frac{1}{W(E)}\{&\psi_1(a,E)\int^{a}_{a}\varphi^T(y,E)F(y)dy \nonumber\\
   &+\varphi_1(a,E)\int^{b}_{a}\psi^T(y,E)F(y)dy\}\label{127},
\end{align}
\begin{align}
  y_2(a,E)=\frac{1}{W(E)}\{&\psi_2(a,E)\int^{a}_{a}\varphi^T(y,E)F(y)dy\nonumber\\
   &+\varphi_2(a,E)\int^{b}_{a} \psi^T(y,E)F(y)dy\}, \label{147}
\end{align}
\begin{align}
  y_1(b,E)=\frac{1}{W(E)}\{&\psi_1(b,E)\int^{b}_{a}\varphi^T(y,E)F(y)dy \nonumber\\
   &+\varphi_1(b,E)\int^{b}_{b}\psi^T(y,E)F(y)dy\}\label{1277},
\end{align}
\begin{align}
  y_2(b,E)=\frac{1}{W(E)}\{&\psi_2(b,E)\int^{b}_{a}\varphi^T(y,E)F(y)dy \nonumber\\
   &+\varphi_2(b,E)\int^{b}_{b} \psi^T(y,E)F(y)dy\}. \label{1477}
\end{align}
Since $\int^a_af(y)dy=0$ and $\int^b_bf(y)dy=0$, where $f(y)$ is the
arbitrary function of $y$,    the equalities (\ref{127}),
(\ref{147}), (\ref{1277}), (\ref{1477})  take the following forms:
\begin{align}
  y_1(a,E)=\frac{1}{W(E)}\{\varphi_1(a,E)\int^{b}_{a}\psi^T(y,E)F(y)dy\}\label{1271},
\end{align}
\begin{align}
  y_2(a,E)=\frac{1}{W(E)}\{\varphi_2(a,E)\int^{b}_{a} \psi^T(y,E)F(y)dy\}, \label{1471}
\end{align}
\begin{align}
  y_1(b,E)=\frac{1}{W(E)}\{\psi_1(b,E)\int^{b}_{a}\varphi^T(y,E)F(y)dy\}\label{12771},
\end{align}
\begin{align}
  y_2(b,E)=\frac{1}{W(E)}\{\psi_2(b,E)\int^{b}_{a}\varphi^T(y,E)F(y)dy\}. \label{14771}
\end{align}
Now, taking into account (\ref{1271}), (\ref{1471}), (\ref{12771}),
(\ref{14771}), we check the conditions (\ref{sy1}), (\ref{sy2}):
\begin{align}
  \label{sy11}
&y_1(a,E)\sin(\alpha)+y_2(a,E)\cos(\alpha)\nonumber\\&=\frac{1}{W(E)}
\{\varphi_1(a,E)\sin{(\alpha)}+\varphi_2(a,E)\cos{(\alpha})\}
\int^{b}_{a}\psi^T(y,E)F(y)dy,
\end{align}
\begin{align}
   \label{sy12}
&y_1(b,E)\sin(\beta)+y_2(b,E)\cos(\beta)\nonumber\\&=\frac{1}{W(E)}
\{\psi_1(b,E)\sin{(\alpha)}+\psi_2(b,E)\cos{(\alpha})\}
\int^{b}_{a}\varphi^T(y,E)F(y)dy.
\end{align}
 Due to the boundary conditions (\ref{5}) for the functions $\psi(x,E)$,
  $\varphi(x,E)$ the relations are
(\ref{sy1}), (\ref{sy2}) are valid. Hence, the matrix $G_0(x,y,E)$
(\ref{G}), is the Green function for the regular boundary problem of
the one-dimensional stationary Dirac equation with the generalized
form of the potential.

The matrix (\ref{G}) can be written in the following form:
\begin{eqnarray}
\label{7} G_0(x,y,E)=(\psi(x) \varphi ^T(y)\Theta(x-y)+
\varphi(x)\psi^T (y)\Theta(y-x))/(W\{\varphi(x),\psi(x)\}),
\end{eqnarray}
where $\Theta(x-y)$, $\Theta(y-x)$ are the Heaviside  step
functions. Thus, we have constructed the Green function of the
initial problem.
\\

\noindent{\bf Example 1} Let us calculate the Green function of the
initial problem for the Dirac (\ref{1}) free particle equation with
the $h_0=i\sigma_2\partial_x+V_0(x)$, $V_0=m\sigma_3$.

We first chose the two linearly independent solutions $\varphi$,
$\psi$ for the free particle case:
\begin{eqnarray}
\label{psi} \psi&=&\left(
           \begin{array}{c}
             \cos{(kx)} \\
             k\sin{(kx)}/(E+m)\\
           \end{array}
         \right),\qquad
  \varphi=\left(
              \begin{array}{c}
                \cos{k(x-1)}\\
                k\sin{k(x-1)}/(E+m)\\
              \end{array}
            \right),\end{eqnarray}\begin{eqnarray}
           k^2&=&{E^2-m^2},\qquad    W\{\varphi,\psi\}=-k\sin{(k)}/(E+m)  \label{varphi} .
\end{eqnarray}

Next we apply the boundary conditions to functions $\psi$, $\varphi$
for the case $a=0$, $b=1$:
\begin{eqnarray}
\label{initialboundaryconditionpsi} \psi\mid_{x=1}&=&\pm\left(
           \begin{array}{c}
             1 \\
             0 \\
           \end{array}
         \right),\qquad \varphi\mid_{x=0}=\pm\left(
              \begin{array}{c}
                1 \\
                0 \\
              \end{array}
            \right).
\end{eqnarray}

From \eqref{psi} it follows that $k=\pi n$, $n\in {\it Z}$.

Also, we find the final expression for the Green function of the
initial regular boundary problem (\ref{s1}), (\ref{s2}) with
$\alpha=\beta=\pm\pi n$:
\begin{eqnarray} \label{Ginitial}
G_0(x,y,E)&=&\frac{A^0\Theta(x-y)+B^0\Theta(y-x)}{W\{\varphi,\psi\}},
\end{eqnarray}
\begin{eqnarray}
\label{A0}\nonumber
  A^0&=&\left(
          \begin{array}{cc}
            \cos{(kx)}\cos{(ky-k)} & k\cos{(kx)}\sin{(ky-k)}/(E+m) \\
            k\sin{(kx)}\cos{(ky-k)}/(E+m) & k^2\sin{(kx)}\sin{(ky-k)}/(E+m)^2\\
          \end{array}
        \right),\\\nonumber
\label{B0}
\end{eqnarray}
$$B^0=(A^0)^T(x\leftrightarrow y).$$

\section{Darboux transformed Green function}
Before constructing Darboux transformed Green function, let us
shortly consider the Darboux transformation method for the Dirac
equation.

The so-called {\it Darbuox transformation operator} $L$ and the
potential $V_1$ of the transformed Dirac Hamiltonian $h_1$ have
respectively the following form:
\begin{eqnarray}
\label{L}  L&=&\partial_x-u_xu^{-1},\\
 \label{V1} V_1&=&V_0+[i\sigma_2,u_xu^{-1}].
\end{eqnarray}
The so-called {\it transformation function} u is $2\times 2$ matrix
consists from two solutions $u=(u_1,u_2)$ of the initial Dirac
equation with $E=\lambda_1,\lambda_2$, where $\lambda_1,$
$\lambda_2$ are the neighboring energy levels.

If the function $\psi$ is solution of initial equation and
$\psi\not=u_1,u_2$, then the function $\tilde{\psi}=L\psi$ is
solution of the transformed Dirac equation
\begin{equation}
\label{111} (h_1(x)-E)\tilde{\psi}(x)=0,\qquad h_1=i\sigma_2+V_1.
\end{equation}

The Green function of the equation (\ref{111}) with the regular
boundary conditions for the  components $\tilde{\Phi}_1(x,E)$,
$\tilde{\Phi}_2(x,E)$
\begin{eqnarray}
 \label{problem}
  \label{s1transformed}
\tilde{\Phi}_1(a,E)\sin(\tilde{\alpha})+\tilde{\Phi}_2(a,E)\cos(\tilde{\alpha})=0,
\end{eqnarray}
\begin{eqnarray}
 \label{s2transformed}
\tilde{\Phi}_1(b,E)\sin(\tilde{\beta})+\tilde{\Phi}_2(b,E)\cos(\tilde{\beta})=0
\end{eqnarray}
of the spinor $\tilde{\Phi}(x,E)=\int_a^bG_1(x,y,E)F(y)dy$ one can
represent in the form:
\begin{eqnarray}
\label{G1} G_1(x,y,E)=(\tilde{\psi}(x) \tilde{\varphi}
^T(y)\Theta(x-y)+ \tilde{\varphi}(x)\tilde{\psi}^T
(y)\Theta(y-x))/(W\{\tilde{\varphi}(x),\tilde{\psi}(x)\}).
\end{eqnarray}

(i) If $\tilde{\alpha}=\alpha$, $\tilde{\beta}=\beta$, we have the
Green function of the Dirac equation with the transformed potential
and the initial regular boundary conditions.

(ii) If $\tilde{\alpha}\not=\alpha$, $\tilde{\beta}\not=\beta$, we
have the Green function of the Dirac equation with the transformed
potential and the modified regular boundary conditions.

Further we consider only the case when $\tilde{\alpha}=\alpha$,
$\tilde{\beta}=\beta$.

Now let us consider the construction of the Darboux transformed
Green function on the basis of the concrete example.

\noindent{\bf Example 2} Let us first chose the transformation
functions as follows:
\begin{eqnarray} \label{matrix}
  u&=& \left(
          \begin{array}{cc}
           1 & \cos{(k_1x)} \\
           0 & k_1\sin{(k_1x)}/(\lambda+m)
          \end{array}
        \right).
\end{eqnarray}

Next, obtain from the free particle case the transformed potential
\begin{eqnarray}
\label{potential}
 V_1&=&-\lambda\sigma_3+k_1\cot{(k_1x)}\sigma_1,
\end{eqnarray}
where \begin{eqnarray} \label{potentialfurnitur} \lambda_1=m,\qquad
\lambda_2=\lambda=\pm\sqrt{m^2+\pi^2}.\end{eqnarray} It can be
checked that only these two transformation matrices (differing from
each other only by sign of $\lambda$) provide the conservation of
the initial regular boundary condition.

\begin{figure}
\centering \includegraphics[height=7cm]{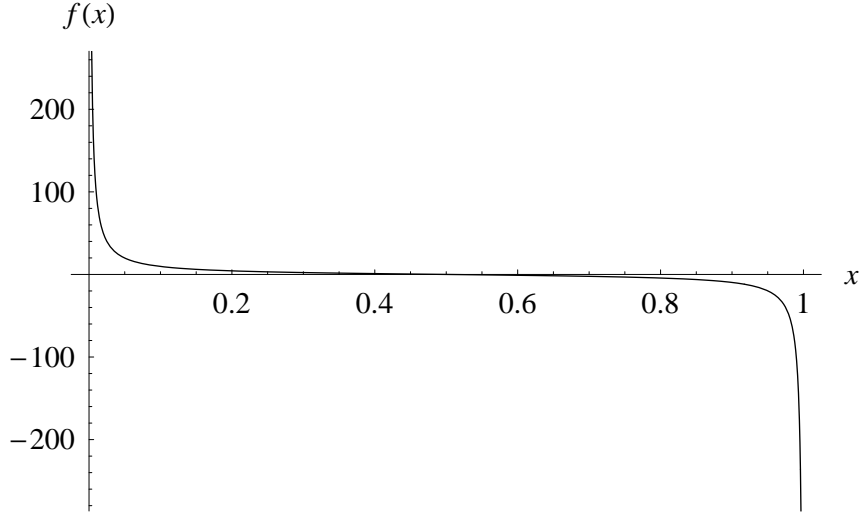} \caption{The
function $f(x)=k_1\cot{(k_1x)}$, $k_1=\pi$} \label{f1}
\end{figure}

Now the Darboux transformed functions $\tilde{\psi}$,
$\tilde{\varphi}$ one can represent in the form:
\begin{eqnarray}
\label{tildapsi}
 \tilde{\psi}&=& \frac{k}{E+m}\left(
 \begin{array}{c}
 (\lambda-E)\sin{kx} \\
 k\cos{kx}-k_1\cot{k_1x}\sin{kx}\\
 \end{array}
 \right),
 \\
 \label{tildavarphi} \tilde{\varphi}&=&\frac{k}{E+m}\left(
 \begin{array}{c}
 (\lambda-E)\sin{k(x-1)} \\
 k\cos{(kx-k)}-k_1\cot{(k_1x)}\sin{(kx-k)} \\
 \end{array}
 \right).
\end{eqnarray}

 Finally, the Green functions
of the Dirac equation with the transformed potential
\eqref{potential} may be given by the formula:
\begin{eqnarray}
\label{Gtransformed}
G_1(x,y,E)&=&\frac{A^{(1)}\Theta(x-y)+B^{(1)}\Theta(y-x)}{W\{\tilde{\varphi},\tilde{\psi}\}},\\
\label{tildaW}W\{\tilde{\varphi},\tilde{\psi}\}&=&(E-\lambda)(E-m)W\{\varphi,\psi\},
\end{eqnarray}
where
\begin{eqnarray}
A^{(1)}=\frac{k^2}{(E+m)^2}\left(\begin{array}{cc} A^{(1)}_{11}&A^{(1)}_{12}\\
A^{(1)}_{21} & A^{(1)}_{22}\end{array}\right),&
&B^{(1)}=(A^{(1)})^T(x\longleftrightarrow y),
\end{eqnarray}
\begin{eqnarray}A^{(1)}_{11}&=&(\lambda-E)^2\sin{(kx)}\sin{(ky-k)}\nonumber,\\\nonumber
A^{(1)}_{21}&=&(\lambda-E)(k\cos{(kx)}-k_1\cot{(k_1x)}\sin{(kx)})\sin{(ky-k)}),\\\nonumber
A^{(1)}_{12}&=&(\lambda-E)\sin{(kx)}(k\cos{(ky-k)}-k_1\cot{(k_1y)}\sin{(ky-k)}),\\\nonumber
A^{(1)}_{22}&=&(k\cos{(kx)}-k_1\cot{(k_1x)}\sin{(kx)})\\\nonumber
&&\times(k\cos{(ky-k)}-k_1\cot{(k_1y)}\sin{(ky-k)}).
\end{eqnarray}
Thus, we have constructed the Green function of the transformed
Dirac equation with the potential \eqref{potential}
and the initial regular boundary conditions.\\

\section{Unabridged trace for the
 difference  of the \\modified and the initial Green functions}

 In this section we calculate the unabridged trace of the
 difference  of  the  transformed and the initial  Green functions.

Firstly, let us consider the  unabridged trace of the transformed
Green function  $tr\int^b_aG_1(x,y,E)|_{x=y}dy$. It can be
represented in the form:
\begin{equation}\label{115}
     tr\int^b_aG_1(x,y,E)|_{x=y}dy=
\frac{tr\int^b_a\tilde{\psi}\tilde{\varphi}^T|_{x=y} dy}
{W\{\tilde{\varphi}\tilde{\psi}\}}.
\end{equation}
It is known that the action of the transformation operator  and the
conjugate transformation operator on the spinors may be written in
the following way \cite{annphys2003v305p151}:
\begin{eqnarray}
  \label{Lpsi1}
   L\psi&=& u\frac{d}{dx}(u^{-1}\psi),\\
  \label{Lpsi1+}
    L^+\tilde{\psi}&=& -(u^{+})^{-1}\frac{d}{dx}(u^{+}\tilde{\psi}).
\end{eqnarray}
Also, accounting these properties, we can rewrite (\ref{115}) as:
\begin{equation}\label{1153}
      tr\int^b_aG_1(x,y,E)|_{x=y}dy=
tr\int^b_a\frac{d}{dx}(u^{-1}\psi)\tilde{\varphi}^Tu|_{x=y}
 dy\biggl/W\{\tilde{\varphi}\tilde{\psi}\}.
\end{equation}

Next let us consider  the unabridged trace of the
 difference  of  the  transformed and the initial  Green functions.
It is obvious \cite{levitan} that the spectral representation for
the initial and the transformed Green functions correspondingly look
like as
\begin{equation}\label{G0}
    G_0(x,y,E)=\sum_{n_0}\frac{\psi_{n_0}(x)\psi^T_{n_0}(y)}{E_n-E},
\end{equation}
\begin{eqnarray}
\label{G0}
    G_1(x,y,E)=\sum_{n_1}\frac{\phi_{n_1}(x)\phi^T_{n_1}(y)}{E_n-E},
\end{eqnarray}
where the functions
\begin{eqnarray}
\label{FI}
\phi_{n_1}=((E_{n_1}-\lambda_1)(E_{n_1}-\lambda_2))^{-1}{L\psi_{n_1}}L,
\end{eqnarray}
form an orthonormal set. The completeness of \eqref{FI} will be
shown in examples 3 and 4.

The construction \begin{eqnarray}
\label{unabrid}tr\int^b_a(G_1(x,y,E)-G_0(x,y,E))|_{x=y}dy\end{eqnarray}
is called the \emph{unabridged trace of the
 difference of  the  transformed and the initial  Green functions}.
 The spectral representation of  \eqref{unabrid} is
 as follows:
\begin{equation}\label{1212127}
     tr\int^b_a(G_1(x,y,E)-G_0(x,y,E))|_{x=y}dy=\sum_{n_1}\frac{1}
     {E_{n_1}-E}-\sum_{n_0}\frac{1}{E_{n_0}-E},
\end{equation}
where $E_{n_{0}}$, $E_{n_{1}}$ are  discrete eigenvalues of  $h_0$
and $h_1$ respectively.

If the set \eqref{FI} is complete then the formula \eqref{unabrid}
one can represent in the form:
\begin{eqnarray}
tr\int^b_a(G_1(x,y,E)-G_0(x,y,E))|_{x=y}dy=\frac{1}{E-\lambda_1}+
\frac{1}{E-\lambda_2}.
\end{eqnarray}

Now let us show an another derivation of the trace formula  for the
Green functions difference. Let us integrate (\ref{1153}) by parts
and apply the trace property
\begin{equation}\label{1152}
tr{AB}=tr{BA}
\end{equation}
to obtain the relation
\begin{equation}\label{1154}
tr\int^b_aG_1(x,y,E)|_{x=y}dy=
\frac{tr\psi\tilde{\varphi}^T|^b_a+\int^b_a tr\psi(L^\dagger
 L\varphi)^T|_{x=y}dy}{W\{\tilde{\varphi}\tilde{\psi}\}}.
    \end{equation}
Due to (\ref{tildaW}) and the factorization property from
\cite{annphys2003v305p151} we obtain:
\begin{equation}\label{1155}
\int^b_atrG_1(x,y,E)|_{x=y}dy=\frac{tr\psi\tilde{\varphi}^T|^b_a}
{W\{\tilde{\varphi}\tilde{\psi}\}}+\int^b_a trG_0(x,y,E)|_{x=y}dy.
\end{equation}
Similarly, we would like to write
\begin{equation}\label{1157}
\int^b_atrG_1(x,y,E)|_{x=y}dy=\frac{tr\tilde{\psi}\varphi^T|^b_a}
{W\{\tilde{\varphi}\tilde{\psi}\}}+\int^b_a trG_0(x,y,E)|_{x=y}dy.
\end{equation}
  Finally, from (\ref{1155}), (\ref{1157}) we find:
\begin{eqnarray}
\label{TIG1}
tr\int^b_a(G_1(x,y,E)-G_0(x,y,E))|_{x=y}dy&=&\frac{1}{W\{\tilde{\varphi},
\tilde{\psi}\}}tr(\psi\tilde{\varphi}^T)\mid^b_a,\\
\label{TIG}
tr\int^b_a(G_1(x,y,E)-G_0(x,y,E))|_{x=y}dy&=&\frac{1}{W\{\tilde{\varphi},
\tilde{\psi}\}}tr(\tilde{\psi}\varphi^T)\mid^b_a.
\end{eqnarray}
Here $a$, $b$ are limits of the integration.
As shown in Appendix, right parts of \eqref{TIG1}, \eqref{TIG} are
equal.

 \noindent{\bf Example 3} In this example we calculate
 \eqref{unabrid} with the help of \eqref{TIG}.

More precisely, take into account \eqref{varphi}, \eqref{tildapsi},
\eqref{tildaW},  we calculate
$\left.tr(\tilde{\psi}\varphi^T)/W\{\tilde{\varphi},\tilde{\psi}\}\right|_a^b,$
where $a=0$, $b=1$:
\begin{eqnarray}
 \left.\frac{tr(\tilde{\psi}\varphi^T)}
 {W\{\tilde{\varphi},\tilde{\psi}\}}\right|_0^1&=
 &\frac{1}{-(E-m)(E-\lambda)\sin(k)}\left\{(\lambda-E)\sin(kx)\cos(kx-k)+\frac{k^2}{E+m}\right.\nonumber\\
 &&\left.\left.\cos(kx)\sin(kx-k)-\frac{k^2}{E+m}\cot(k_1x)\sin(kx)\sin(kx-k)\right\}\right|_0^1.\nonumber
\end{eqnarray}
Using  the  L'Hospital's rule for evaluating of indeterminations, we
obtain
\begin{eqnarray}
 \label{tracedG}
 \left.\frac{tr(\tilde{\psi}\varphi^T)}
 {W\{\tilde{\varphi},\tilde{\psi}\}}\right|_0^1&=
 &\frac{1}{E-m}+\frac{1}{E-\lambda}.
\end{eqnarray} Due to  \eqref{TIG} we get:
\begin{eqnarray}
tr\int^1_0(G_1(x,y,E)-G_0(x,y,E))|_{x=y}dy=\frac{1}{E-\lambda_1}+\frac{1}{E-\lambda_2},\end{eqnarray}
where $\lambda_1=m$, $\lambda_2=\lambda$.
This result indicate that the spectrum of the transformed Dirac
Hamiltonian differs from the spectrum of the initial Dirac operator.
The spectral lines $E=\lambda,m$ disappear from the transformed
spectrum and  the set of functions $\phi_{n_1}$ is
complete.\\

 \noindent{\bf Example 4} In this example we calculate
 \eqref{unabrid} in a direct way.

First, we consider
\begin{eqnarray}
\label{trG0}
trG_0(x,x,E)=-\frac{E}{k}\cot(k)-\frac{m\cos(2kx-k)}{k\sin(kx)}.
\end{eqnarray}
Next, we integrate the expression (\ref{trG0}) \begin{eqnarray}
\label{intrG0}
tr\int_0^1G_0(x,x,E)dx=-\frac{E}{k}\cot{k}-\frac{m}{k^2},
\end{eqnarray}
$$k=\pi n,\quad \cot(k)-k^{-1}=\Sigma_{n=1}^{\infty}2k/(E^2-E_n^2),$$
\begin{eqnarray} \label{intrG0}
tr\int_0^1G_0(x,x,E)dx=-\Sigma_{n=1}(E+E_n)^{-1}-\Sigma_{n=1}^{\infty}(E-E_n)^{-1}-(E-m)^{-1}.
\end{eqnarray}

Similarly, we calculate \begin{eqnarray} \label{inttrG1}
tr\int_0^1G_1(x,x,E)=
-\Sigma_{n=1}(E+E_n)^{-1}-\Sigma_{n=1}^{\infty}(E-E_n)^{-1}+(E-\lambda)^{-1}
\end{eqnarray}
and  obtain that
\begin{eqnarray}
tr\int{G_1(x,x,E)-G_0(x,x,E)}dx&=&
\frac{1}{E-\lambda}+\frac{1}{E-m}.
\end{eqnarray}
Thus, we can conclude that the set of functions $\phi_{n_1}$ is
complete.

\section{Conclusion}
 In this paper, we have
studied the Darboux transformation of the Green functions of the
regular boundary problem corresponding to the initial and the
transformed potentials of the one-dimensional Dirac equation for the
case of the Dirac Hamiltonians with discrete spectrum. The main
results of the paper are the construction of the Darboux transformed
Green function with initial regular boundary conditions  and the
formulae for an unabridged trace \eqref{TIG1}, \eqref{TIG}. For the
checking of formulae \eqref{TIG1} and \eqref{TIG} we consider the
unabridged trace of difference between transformed and initial Green
functions by usual possible and by formula \eqref{TIG1}, both
results are equal.  The all results of this paper are studied only
for discrete spectrum and regular boundary problem. An interesting
question for the future is generation of these results  to the case
of continuous spectrum of the Dirac equation and the case of problem
on the real line  and a  half-line. We believe that these problem
will investigate in a separate publication.

\section*{Acknowledgments}

The author is grateful to  Dr. D. Antonov for helpful comments of
the given work. I would like also the express my thanks to Prof. B.
G. Bagrov for useful discussions. This work was supported in part by
the ``Dynasty'' Fund and Moscow International Center of Fundamental
Physics.

\section*{Appendix. Equality of traces}
In this Appendix, taking into account \eqref{psi}, \eqref{varphi},
\eqref{tildapsi}, \eqref{tildavarphi}, we show the equivalence of
expressions for right parts of \eqref{TIG1}, \eqref{TIG}
\begin{eqnarray}
\label{tracepsi}
  tr(\psi\tilde{\varphi}^T)\mid_a^b&=&(\psi_1\tilde{\varphi}_1+\psi_2\tilde{\varphi}_2)\mid_a^b=\psi_1(b)\tilde{\varphi}_1(b)\nonumber\\
  &&+\psi_2(b)\tilde{\varphi}_2(b)-\psi_1(a)\tilde{\varphi}_1(a)-\psi_2(a)\tilde{\varphi}_2(a),\\
\label{tracetildapsi}
tr(\tilde{\psi}\varphi^T)\mid_a^b&=&(\tilde{\psi}_1\varphi_1+\tilde{\psi}_2\varphi_2)\mid_a^b=\tilde{\psi}_1(b)\varphi_1(b)\nonumber\\
&&+\tilde{\psi}_2(b)\varphi_2(b)-\tilde{\psi}_1(a)\varphi_1(a)-\tilde{\psi}_2(a)\varphi_2(a),
\end{eqnarray}
where $a=0$, $b=1$.

In the explicit form \eqref{tracepsi}, \eqref{tracetildapsi} look
like  as follows:
\begin{eqnarray}
\label{tracepsifunction}
  tr(\psi\tilde{\varphi}^T)\mid_0^1&=&\nonumber
\{(\lambda-E)\cos(kx)\sin(kx-k)+(E-m)\sin(kx)\cos(kx-k)\\
&&-\frac{kk_1}{E+m}\cot(k_1x)\sin(kx)\sin(kx-k)\}\mid_0^1,
\end{eqnarray}
\begin{eqnarray}
\label{tracetildapsifunction}
  tr(\tilde{\psi}\varphi^T)\mid_0^1&=&\nonumber
\{(\lambda-E)\sin(kx)\cos(kx-k)+(E-m)\cos(kx)\sin(kx-k)\\
&&-\frac{kk_1}{E+m}\cot(k_1x)\sin(kx)\sin(kx-k)\}\mid_0^1.
\end{eqnarray}
Since\begin{eqnarray}
            tr(\psi\tilde{\varphi}^T)\mid_0^1-tr(\tilde{\psi}\varphi^T)\mid_0^1&=&
((\lambda-E)\sin(-k)+(E-m)\sin(k))\mid_0^1=0
\end{eqnarray}
we finally  obtain \begin{eqnarray} \label{eqvevalenceoftraces}
   tr(\psi{\tilde{\varphi}}^T)\mid_0^1 = tr(\tilde{\psi}\varphi^T)\mid_0^1.
     \end{eqnarray}


\begin{thebibliography}{0}
\bibitem{Darbu} G. Darboux, {\it Compt. Rend. Acad. Sci.}, Paris {\bf 94} 1343 (1882); ibid. {\bf 94} 1456
(1882).

\bibitem{Rosu}H. C. Rosu, {\it Short Survey of Darboux Transformations},~arXiv:quant-ph/980956.


\bibitem{Axel1} A. Schulze-Halberg,
{\it Int. J. Mod. Phys.,} A {\bf 22} 1735 (2007).

\bibitem{Axel2} A. Schulze-Halberg,
{\it Int. J. Mod. Phys., } A {\bf 21} 4853 (2006).

\bibitem{Song2} D.-Y. Song and J. K. Klauder,
{\it J. Phys.}, A {\bf 38} 5831 (2005).

\bibitem{Matv} V. B. Matveev and M. A. Salle,
 {\it Darboux Transformations and Solitons} (Berlin: Springer, 1991).

\bibitem{Suzko} A. A. Suzko,
{\it Int. J. Mod. Phys.}, A {\bf 12} 277 (1997).

\bibitem{Pozdeeva} E. O. Pozdeeva,
 {\it J.  Surf. Invest.}, {\bf 3} 66 (2007).

\bibitem{Park} Q.-H. Park and H. J. Shin, M. A.,
 {\it Physica}, D {\bf 157} 1 (2001).

\bibitem{Abl} M. J. Ablowitz and H. Segur,
 {\it Solitons and the Inverse  Scattering Transform} (Philadelpia: SIAM, 1981).

\bibitem{Witten} E. Witten, {\it Nucl. Phys.}, B {\bf 185} 513 (1981).

\bibitem{BAGROVECHAR}  V. G. Bagrov and  B. F. Samsonov, {\it Phys. Part. Nucl.}, {\bf
28} 374   (1997).

\bibitem{Sukumar1985}  C. V. Sukumar  {\it J. Phys.}, A {\bf 18}
L57 (1985).

\bibitem{BagrovSamsonov} V. G. Bagrov and  B. F. Samsonov,
{\it Theor. Math. Phys.}, {\bf 104} 356 (1995).


\bibitem{Gomez} D. Gomez-Ullate, N. Kamran and R. Milson,  {\it J. Phys.}, A {\bf 37}
10065 (2004).

\bibitem{annphys2003v305p151} L. M. Nieto, A. A. Pecheritsin  and
 B. F. Samsonov,  {\it Ann. Phys.}, {\bf 305} 151 (2003).

\bibitem{Eurjphys}  B. F. Samsonov, A. A. Pecheritsin, E. O. Pozdeeva
et al., {\it Eur. J. Phys.}, {\bf 24} 435 (2003).

\bibitem{BAGROV} V. G. Bagrov,   A. A. Pecheritsin, E. O. Pozdeeva
et al., {\it Commun.  Nonlin. Scien.}, {\bf 9} 13 (2004).

\bibitem{Sukumar} C. V. Sukumar,
{\it J. Phys.}, A {\bf 37} 10287 (2004).

\bibitem{Samsonov} B. F.Samsonov, C. V. Sukumar and A. M. Pupasov,
 {\it  J. Phys.}, A {\bf 38} 7557 (2005).

 \bibitem{Pupasov} A. M. Pupasov and B. F. Samsonov,  {\it Russ. Phys. J.}, {\bf 48} 1020 (2005).


\bibitem{Anderson} A. Anderson,  {\it Phys. Rev.}, D {\bf 37} 536
(1988).
 \bibitem{Axel4} A. Schulze-Halberg, {\it Commun. Theor. Phys.}, {\bf 41} 723 (2004).

 \bibitem{Deber} N. Debergh, A. A. Percheritsin, B. F. Samsonov et
 al.,
{\it J. Phys.}, A {\bf 35} 3279 (2002).

\bibitem{levitan} B. M. Levitan and I. S. Sargsjan, {\it Sturm-Liouville
and Dirac Operator} (Kluwer Academic Publishers, 1991).

\bibitem{thaler} B. Thaller,
 {\it The Dirac Equation} (Berlin: Springer, 1992).

\bibitem{107}  P. F. Filchakov,  {\it Mathematical handbook}
(Kiev: Naukova Dumka, 1973).
\end{thebibliography}
\end{document}